\documentclass{proc}
\usepackage{amsfonts}
\usepackage{amsmath}
\usepackage{graphicx}
\usepackage{footmisc} 

\usepackage{balance}
\usepackage[square, comma, sort&compress, numbers]{natbib}
\usepackage{subfigure}
\usepackage{booktabs}
\usepackage{multirow}
\usepackage{paralist}
\usepackage[all]{xy}

\usepackage[hyphens]{url}
\usepackage{hyperref}
\hypersetup{
    colorlinks,
    citecolor=black,
    filecolor=black,
    linkcolor=black,
    urlcolor=black
}

\newcommand{\spara}[1]{\smallskip\noindent{\bf #1}}

\usepackage{array}
\newcolumntype{L}[1]{>{\raggedright\let\newline\\\arraybackslash\hspace{0pt}}p{#1}}
\newcolumntype{C}[1]{>{\centering\let\newline\\\arraybackslash\hspace{0pt}}p{#1}}
\newcolumntype{R}[1]{>{\raggedleft\let\newline\\\arraybackslash\hspace{0pt}}p{#1}}

\usepackage{verbatim}

\begin{document}

\title{Understanding Types of Users on Twitter}
\author{Muhammad Moeen Uddin$^{1}$, Muhammad Imran$^{2}$, Hassan Sajjad$^{2}$ \\
Lahore University of Management Sciences, Lahore, Pakistan$^{1}$\\
Qatar Computing Research Institute, Doha, Qatar$^{2}$\\
moeen.din@apparelco.com, mimran@qf.org.qa, hsajjad@qf.org.qa}
\maketitle

\begin{abstract}
People use microblogging platforms like Twitter to involve with other users for a wide range of interests and practices. Twitter profiles run by different types of users such as humans, bots, spammers, businesses and professionals. This research work identifies six broad classes of Twitter users, and employs a supervised machine learning approach which uses a comprehensive set of features to classify users into the identified classes. For this purpose, we exploit users' profile and tweeting behavior information. We evaluate our approach by performing 10-fold cross validation using manually annotated 716 different Twitter profiles. High classification accuracy (measured using AUC, and precision, recall) reveals the significance of the proposed approach.
\end{abstract}

\section{Introduction}
\label{intro}
Microblogging platforms have become an easy and fast way to share and consume information of interest on the Web in real-time. For instance, in recent years, Twitter\footnote{http://twitter.com/} has emerged as an important source of real-time information exchange platform. It has empowered citizens, companies, marketers to act as content generators, that is, people share information about what they experience, eyewitness, and observe about topics from a wide range of fields such as epidemics~\cite{culotta2010towards}, disasters~\cite{imran2013practical}, elections~\cite{tumasjan2010predicting} and more.

To consume information, Twitter users follow other users who they think can provide useful information of their interest. Information shared on Twitter in the form of short text messages (``tweets") immediately propagated to followers, and implicitly starts a one-way conversation, which is also known as social interaction \cite{fischer2011social}. Often such conversations turn in two-way when followers reply back. Further spread of the information happens when followers post the received information to their followers (i.e., re-tweeting).

We believe that social interaction on social media has a resemblance to social interaction that one practices in daily routine. For instance, companies leverage insights from social media information to better market to its customers and increase sales. In this case, companies always seek to gain more in-depth information of their customers for better understanding and to improve interaction with them despite it is one-to-one, through a phone call, or on social media.



Moreover, understanding the types of users on social media is important for many reasons. To name a few, for example this includes detecting bots or spam users~\cite{benevenuto2010detecting}, recommending friends (e.g., potential users to follow on twitter)~\cite{hannon2010recommending}, finding credible information and users~\cite{castillo2011information}, for example, to receive trusted analysis or feedback of products or to ask questions to fulfill information needs~\cite{paul2011twitter}, and so on. Moreover, prior knowledge of audience helps companies, marketers, NGOs to classify their followers into different categories (e.g., personal, professional, bots, etc.) to have an effective and targeted interaction with them.

In recent years, Twitter has been extensively used in a number of research studies that analyze and process mainly tweets content using different natural language processing (NLP) techniques to differentiate Twitter users~\cite{Pennacchiotti:ICWSM2011}. Moreover, many studies focus on aspects like, who follows whom, who is in which list, etc. However, understanding the types of twitter users using their tweeting behavior or more importantly what their profile information reflects, is an aspect which is broadly overlooked. Twitter profiles provide useful information, furthermore determining various behavioral aspects of users on Twitter such as how often they post, re-tweet, or reply could provide significant insights about users.

In this paper, we study Twitter from a different perspective, that is, we categorize Twitter users into different classes by exploiting their profiles and tweeting behavior information. Based on our manual investigation of randomly selected 716 Twitter profiles, we identify six broad classes of Twitter users. Extensive set of comprehensive features were learned during the manual analysis phase, which are then used to train a machine learning classifier to automatically classify Twitter profiles. Furthermore, validation of our hypothesis is conducted by performing 10-fold cross validation of the trained classifier. Finally, we claim that the proposed approach can effectively classify followers of a given Twitter profile into the proposed classes.

Rest of the paper is organized as follows. In the next section, we discuss Twitter profile, user behavior and content specific information. Based on that, we present six user classes. Section \ref{sec:methodology} describes our research framework. In section \ref{sec:models}, we introduce our features set and model used for learning. Section \ref{sec:models} reports results of our experimentation and section \ref{sec:relatedWork} summarizes the related work. Finally, we conclude the paper in section \ref{sec:conclusion}.

\section{Profile and Tweeting Behavior Specific Information}
\label{sec:information}
Twitter users can be analyzed based on their profiles, posts, and tweeting behavior. Users' profiles exhibit an extensive set of informational pieces, users' posts represent rich content (i.e., tweets) often used to perform NLP based analysis, and users' tweeting behavior represents different aspects related to a user's interaction with the platform as well as with other users (e.g., followers). In figure 1 we show a partial view of the information that can be obtained from Twitter about a user. It shows a meta-data part (i.e., profile specific information, followers, and friends), and a content part (i.e., tweets).
To classify Twitter users into different classes, we exploit users' profile and their tweeting behavior. Following subsections expand both aspects in detail.

\subsection{Profile specific information} Users on Twitter can be anyone. These users can be classified into two broad categories, that are, (i) \emph{real-users}, (ii) \emph{digital-actors}. Real-users represent human-beings (e.g., home users, business users, or professional users), and digital-actors represent automated computer programs (e.g., bots, online services, etc). Both types of user built their profiles on Twitter by specifying information such as \emph{name, website, description, bio,} etc. Other information such as \emph{created\_at, status\_count, listed\_count} that a twitter profile contains automatically provided or manipulated by Twitter platform and it tends to change over time (e.g., number of followers change over time, listed\_count change over time). 

In general, tweets posted by users are publicly available and are followed by subscribers called \emph{followers}. Users who share particular interests are included in one's reading \emph{list}. A profile's \emph{listed count} is the number of users whose reading lists contain the profile's tweets.

\subsection{Information based on tweeting \\ behavior}
We define tweeting behavior as a collective measure of a user interactions on Twitter; that includes number of tweets he/she posted, number of re-tweets, and replies. 

We consider re-tweets as a form of endorsement to particular tweets. Especially, re-tweets to different tweets of different users represent a more natural behavior, and a reply as a more concerned opinion on a topic. We consider such behaviors closer to the behavior that we expect from real-users, whereas, on the other hand constantly re-tweeting tweets of a specific set of users, less replying, and tweeting following a fixed pattern show the behavior of digital-actors.

\begin{figure}[t]
\includegraphics[width=\columnwidth]{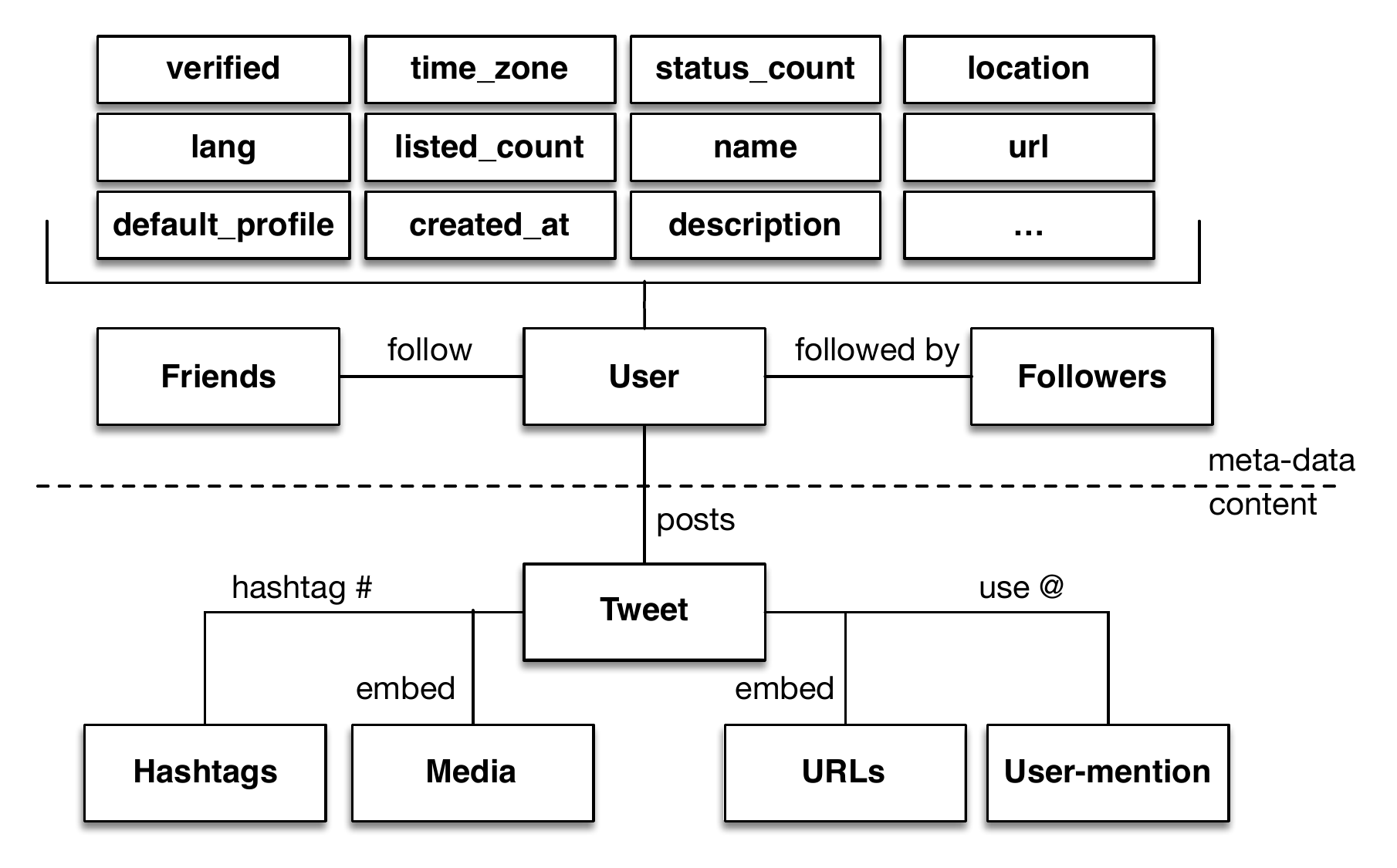}
  \caption{Twitter infographic}
  \label{fig:twitter}
\end{figure}

\subsection{User Classes}
Based on the information provided by Twitter profiles and users' tweeting behavior as explained above, we categorize twitter users into six different classes, three of them are of type real-users and three belongs to digital actors. They are described as follows:

\begin{itemize}
\item \emph{Personal users}: We consider personal users as casual home users who create their Twitter profile may be for fun, learning, or to acquire news, etc. These users neither strongly advocate any type of business or product, nor their profiles are affiliated with any organization. Generally, they have a personal profile and show a low to mild behavior in their social interaction.

\item \emph{Professional users}: They are home users with professional intent on Twitter. They share useful information about specific topics and involve in healthy discussion related to their area of interest and expertise. Professional users tend to be highly interactive, they follow many, and also followed by many. 


\item \emph{Business users}: Business users are different than personal/professional users in that they follow a marketing and business agenda on Twitter. The profile description strongly depicts their motive and a similar behavior can be observed in their tweeting behavior. Frequent tweeting, less interaction are two key factors that distinguish business users from both personal and professional users.

\end{itemize}

The next three classes of users are of type digital actors. Common features that these users share include highly frequent tweeting, no or less interactivity, and mostly their followers either increase (e.g., in case of feed/news users) or decrease (e.g., in case of spam users) over time. 

\begin{itemize}
\item \emph{Spam users}: Spammers mostly post malicious tweets at high rate. Mostly, automated computer programs (bots) run behind a spam profile, and randomly follow users, expecting a few users to follow back. Sometimes, personal users can also behave as a spammer, but often they do not get caught because their spamming behavior do not follow a pattern, which can be easily seen in case of an automated spam profile. Moreover, followers of a spam users decrease over time. 

\item \emph{Feed/news}: These profile types represent automated services that post tweets with information taken from news websites such as CNN, BBC, etc. or from different RSS feeds. Like spammers, often tweets posting by these profiles is controlled by bots. The key difference between spammers and these profiles is the increase in followers count over time. Moreover, these users are not interactive at all (i.e., zero replies). 

\item \emph{Viral/Marketing Services}: Viral marketing, or advertising refer to the marketing techniques that marketers use with the help of technologies/social networks to increase their brand awareness, sale, or to achieve other marketing objectives. People use a viral process, which is an advance type of a bot (i.e., an intelligent bot that spreads information also produces fake likes, followers, etc.), to accomplish their marketing tasks.

\end{itemize}




\section{Methodology}
\label{sec:methodology}
We have presented different classes and their characteristics in which Twitter users can be classified. In this section, we describe our approach regarding data collection, annotation. Moreover, we show results of our manual annotation that help us choose more prominent features to be used for the automatic classification of users. Following subsections describe them in detail.

\begin{table}[htb]%
\caption{Keywords/phrases and trends used for users' profiles collection.\label{table:1}}
\centering
\medskip
\begin{tabular}{l | r r}
\toprule
{\bf Keyphrase} & {\bf Users} & {\bf Tweets} \\
\midrule
Cars & 147 & 14,700 \\
Coffee & 144 & 14,300 \\
News & 93 & 9,300 \\
Repairs & 50 & 5,000 \\
Web application & 14 & 1,400 \\
\midrule
{\bf Trend} & {\bf Users} & {\bf Tweets} \\
\midrule
Callme & 169 & 16,900 \\
WeekofBTR & 79 & 7,900 \\
Ten\~Af\^A-aQueTwittearlo& 20 & 2,000 \\
QueTwittearlo & 20 & 2,000 \\
\midrule
{\bf Total} & {\bf 716} & {\bf 71,600} \\
\bottomrule
\end{tabular}
\label{tab:corpusCollection}
\end{table}

\subsection{Data Collection and Annotation}

\spara{Data collection:} We collect user profiles over a span of one week by using Twitter streaming API\footnote{\url{https://dev.twitter.com/docs/streaming-api}}. To this end, we used Java-based open-source code available on Git by the AIDR platform~\cite{imran2014aidr}. In order to cover diverse user types in terms of intent, interest, background and behavior in our dataset, we select users by using different keywords/phrases, and world-wide trending topics on twitter. We randomly choose 900 profiles, out of which 184 were dropped because we found some profile information was missing. The final dataset contains 716 profiles in total, in which 448 profiles are collected based on key-phrases and 268 profiles are collected based on trends. 

Table \ref{tab:corpusCollection} shows keywords/phrases and trends that we used for the collection and the distribution of profiles among them. In addition to the collected profiles, we downloaded last hundred tweets of each profile.

Based on the collected dataset, in Table 2 we show a few users' profile specific statistics, and in Table 3 we show some accumulative measures of users.

\begin{table}[htb]%
\caption{Users' profile specific measures.\label{tab:collection-user-stats}}
\centering
\medskip
\begin{tabular}{l | r r}
\toprule
{\bf Profile info.} & {\bf Users} & {\bf Percent} \\
\midrule
Verified & 170  & 23\% \\
Zero favorite &  124 & 17\% \\
At least one reply & 593 & 82\% \\
no promotion & 227 & 32\% \\
$>$ 10 promotions & 126 & 18\% \\
$>$ 5 re-tweeted status & 405 &  57\% \\
$>$ 20 URLs &  252 & 36\% \\
$>$ one hour of age & 39 & 5\% \\
$>$ 10 replies & 406 & 57\% \\
100 tweets in one hour & 253 & 35\% \\
\bottomrule
\end{tabular}
\end{table}

\spara{Data annotation:} To annotate the collected dataset, we implement a simple computer program (using Java language) to measure important profile specific statistics. For example, the most prominent ones are shown in Table 2, and in Table 3. Next, based on the characteristics described in section \ref{sec:information}, we manually classified the profiles into the six classes that are proposed in section \ref{sec:information}. For cases where some user qualifies to be categorized in more than two classes, we cross checked their identity by visiting URL provided in the description of their profiles.

\begin{table}[htb]%
\caption{Accumulative measures specific to users' behavior.\label{tab:collection-stats}}
\centering
\medskip
\begin{tabular}{l | r r}
\toprule
{\bf Statistic type} & {\bf Occurrences} & {\bf Percent} \\
\midrule
Tagging other users & 55,429 & 77\% \\
Embedded links & 22,663 & 31\% \\
Retweets & 10,710 & 14\% \\
hashtags & 22,273 & 31\% \\
\bottomrule
\end{tabular}
\end{table}

\spara{Results of manual annotation:} Table 4 shows distribution of the users into the six classes. Figure 2 shows some insights from our manual annotation results. From the results, it can be clearly observed that \emph{business users} are more interactive than personal and professional. Whereas, the percentage of endorsements (i.e., retweets of a particular tweet) of \emph{professional users} prominently are more among the users in real-users category. We also note that \emph{spam} and \emph{feed/news} users do more re-tweets, and almost zero replies are identified in their cases. Generally, among all categories we found a large proportion of users are not verified.

\begin{table}[htb]%
\caption{Results of manual classification of users into six classes.\label{tab:manual-distribution}}
\medskip
\centering
\begin{tabular}{l | r}
\toprule
{\bf Class} & {\bf \# of Users}  \\
\midrule
Personal & 19 \\
Professional & 399 \\
Business & 157 \\
Spam & 49 \\
Feed/news & 51 \\
Viral & 41 \\
\midrule
{\bf Total} & {\bf 716} \\
\bottomrule
\end{tabular}
\end{table}



\includegraphics[width=\columnwidth]{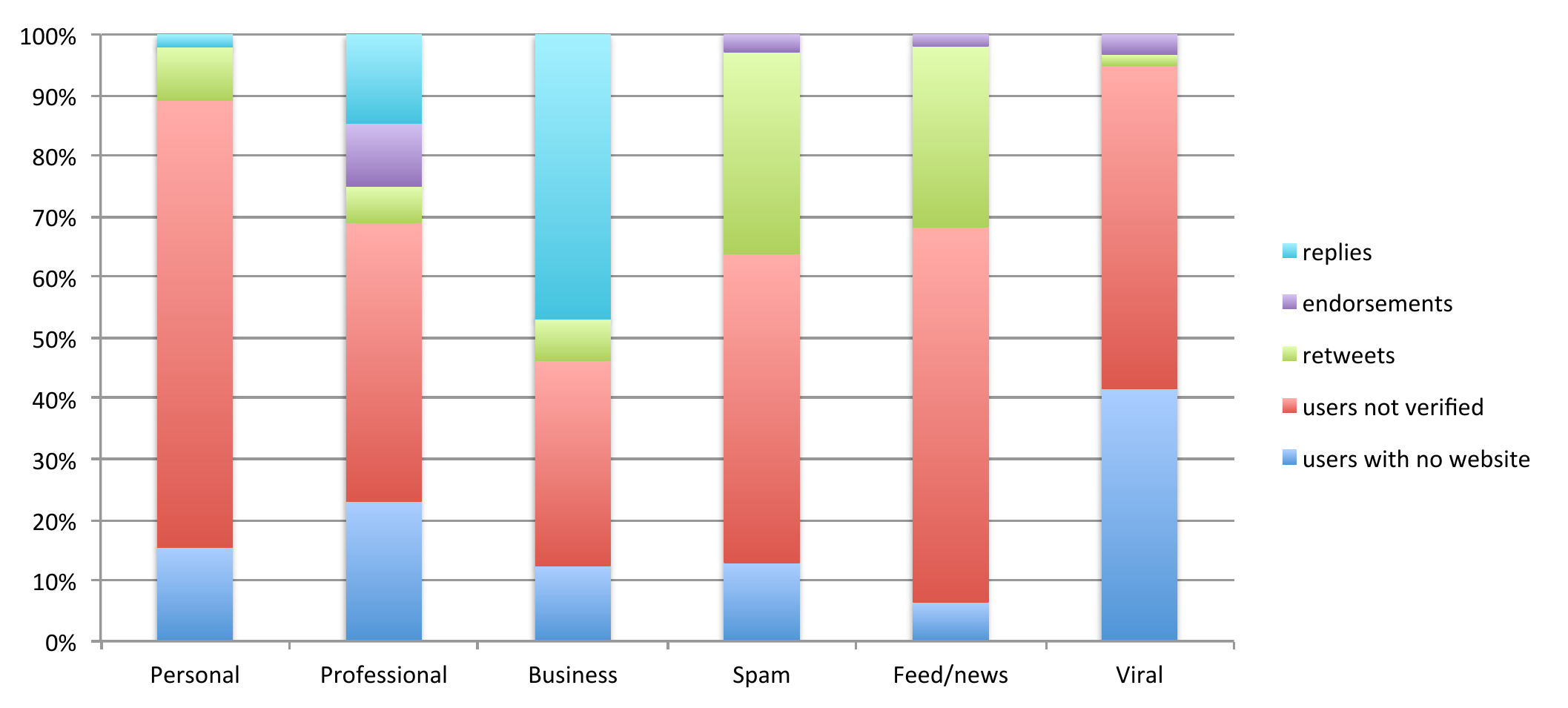}
\label{fig:manual-results}
\hypertarget{figure2}{Figure-2. Results of manual annotation of the collected dataset based on five features} 



\section{Automatic Classification of \\ Twitter Users}
\label{sec:models}
Results obtained from the manual annotation clearly show that our dataset has unbalanced class representation and lack of representation
can produce overfitting for learner for those particular classes. For this purpose, we need a classification algorithm that can clearly discriminate among the classes but also avoids overfitting. As reported in \cite{Nallapati:2004}, discriminative models are preferred over generative models. They also tend to have a lower asymptotic error as the training set size is increased. Simple regression based algorithms can also give good results for some classes but some experiments show poor classification accuracy for minority classes.

When dealing with unbalanced class distributions, discriminative algorithms such as support vector machine (SVM), which is a well known machine learning technique, maximizes the classification accuracy result in case of trivial classification by ignoring minority classes. To avoid this situation, we choose random forest classification technique with bagging approach for our automatic classification purposes.

\subsection{Selected feature set}

To classify Twitter profiles into the defined classes, we choose 17 features as summarized below. These features include a few trivial ones, which can be easily obtained from profiles, for example, statistical features like \# of tweets, \# of replies, total URLs, etc. However, some of the selected features are derived like \emph{Std hashtag, Std URLs, Collective influence, etc.} 

\begin{enumerate}
\item \emph{Favorites count:} represents the number of tweets of a user which were favorited by others.

\item \emph{Verified:} specifies whether an account is verified or not.

\item \emph{Plain statuses:} represents the number of plain statuses that are posted without hashtags, URLs, or mentions of other users.

\item \emph{Replies received:} represents the number of statuses that received replies from other users.

\item \emph{Replies given:} specifies the number of responses/replies given on other users' statuses.

\item \emph{Retweets:} represents the number of retweets posted by a user using other users' tweets.

\item  \emph{Mentions:} represents the number of mentions found in the tweets posted by a user.

\item \emph{Total URLs:}  represents total number of URLs used in tweets. 

\item \emph{Total hashtags: } represents total number of hashtags used in tweets.

\item \emph{Promotion score:} represents the edit distance between the expanded$\_$url and the user profile name. Comparing promotion score of users of different classes gives interesting insights about self branding or self promotion.

\item \emph{Life time:} specifies the life time of a user, measured using the time mentioned in a profile's created\_at field and time of the last tweet.

\item \emph{Tweet spread/influence:}  represents how influential a given tweet is. It is calculated using (retweets count a tweet received $/$ total time taken for first 100 retweets.

\item \emph{Std URLs:} represents the standard deviation of URLs that a user embedded in his tweets.

\item \emph{Std hashtags} represents the standard deviation of hashtags that a user used in his tweets.


\item \emph{User collective activeness:}  represents how active a user is. It is calculated using the total number of posted tweets, number of following, and number of lists for a given time period. Time period can be a week, month, or several months.

\item \emph{Degree of inclination:} shows how a user urge to act or feel in a particular way. We measure it as the Harmonic mean of personal statuses and retweeted status.

\item \emph{Collective influence:}  represents a collective measure of a user's influence. It is measured using sum of followers, user's listed count, and favorite count.
\end{enumerate}

\subsection{Evaluation}
To evaluate our approach, we perform classification of manually annotated data using the above mentioned features. We perform a 10-fold cross validation taking 9/10 of the data as training and 1/10 as test data. Each fold comprises of roughly 71 users. We normalize the value of features and train a classifier. We classify them using bagging with random forest classifier. The results are shown in Table 5. 

In this setup, we measure classification accuracy using precision, recall, F-measure, and AUC measures. We mainly rely on AUC measure, as it is considered to be more reliable than others. As the prevalence of \emph{professional}, and  \emph{business classes} is more, that is the reason we can clearly observe high classification accuracy for these classes. Overall, the classifier we learn performed well in all cases except \emph{feed/news}.


\begin{table}[htb]%
\caption{Cross validation results using random forest classification with bagging technique.\label{tab:resultCross}}
\centering
\medskip
\footnotesize
\begin{tabular}{l | c c c c }
\toprule
Class & Precision & Recall & F-Measure & AUC  \\
\midrule
Personal users & 0.692 & 0.579 &  0.629 & 0.962  \\
Professional users & 0.872 & 0.942  & 0.906 & 0.970 \\
Business users & 0.895 & 0.933 & 0.914 & 0.990  \\
Spam users & 0.532 & 0.510 & 0.521 & 0.936 \\
Feed/news & 0.512 & 0.431 & 0.468  & 0.934\\
Viral/marketing & 0.711 & 0.780 & 0.744 & 0.970 \\
\bottomrule
\end{tabular}
\end{table}


\label{sec:evaluation}
\section{Related Work}
User classification on twitter is a non-trivial task. Most of the related work includes different features to classify users and there are various dimensions in which users have be classified.
For instace, in \cite{Pennacchiotti:2011}, and \cite{Rao:2010} authors use linguistic, profile, and social network features to classify users into political affiliations. In \cite{Dokoohaki:2012}, authors exploit the ``list" feature of the twitter and classify elite users as celebrities, bloggers, and representatives of media outlets and other formal organizations. They have snowball sampled from lists. Whereas, in \cite{Tinati:2012} authors classify users based on their communicator roles -- amplifier, curator, idea starter and commentator. The features that this work considered include influence factors and retweeting feature.


We consider work presented in \cite{Chu:2012} is more closer to our approach, however, they classify users only to detect bots, whereas we consider a more broad set of users. Their work observe tweeting behavior, tweet content, and account properties to identify features that are different for human, bot and cyborgs. Their classification method consists of entropy-based component, spam detector, account properties component and a decision maker. In \cite{Tao:2011} authors classify users on the basis of hashtags, topic of interests and entity based profiles. Their objective is to recommend fresh content to the user based on their class. Work presented in \cite{Cha10measuringuser} measures the dynamics of user influence based on in-degree, retweets and mentions. Our work is different from the previous work as we categorize users based on their personal attributes that they mention in profiles as well as their tweeting behavior. Moreover, we focus on a broad set of user categories that none of the related work focused on. 

\label{sec:relatedWork}
\section{Conclusion \& Future Work}
Twitter is a famous microblogging platform used by companies, businesses, professionals, and also by home users in their daily routine to disseminate information Online in real-time. Twitter users exhibit different characteristics that distinguish one user from others. Understanding Twitter users is important for many reasons such as for companies to plan their marketing campaigns differently for different types of users. 

In this paper, we study Twitter to classify its users into different classes. We identified six different classes based on different characteristics that we observed by studying almost 716 Twitter profiles. Moreover, we performed automatic classification of Twitter users employing supervised machine learning technique by using most prominent features that can effectively be used for classification of Twitter users. High classification accuracy of our experiments show the significance of our approach. 

Tweeting behavior of Twitter users change over time. Monitoring changes related to a user interaction with the platform as well as with other users could significantly reveal more insights and ultimately strengthen the classification accuracy we achieved in our experimentation. We leave this aspect as a potential future work to be explored.




\label{sec:conclusion}

\bibliographystyle{IEEE}
\balance
\bibliography{paper}

\end{document}